\documentstyle[amssymb,aps,prl,epsf,floats,amsbsy]{revtex}

\def\boldk{{\bf k}}
\def\boldR{{\bf R}}
\def\bold0{{\bf 0}}

\def\Schrodinger{Schr\"odinger }

\begin{document}
\twocolumn[\hsize\textwidth\columnwidth\hsize\csname@twocolumnfalse%
\endcsname

\title{Systematic first-principles study of impurity hybridization in NiAl}
\author{David Djajaputra and Bernard R. Cooper}
\address{
Department of Physics, West Virginia University, P.O. Box 6315, Morgantown, 
WV 26506-6315}

\date{\today}

\maketitle

\begin{abstract}
We have performed a systematic first-principles computational study of the 
effects of impurity atoms (boron, carbon, nitrogen, oxygen, silicon, 
phosporus, and sulfur) on the orbital hybridization and bonding 
properties in the intermetallic alloy NiAl using a full-potential linear
muffin-tin orbital method. The matrix elements in momentum space were used
to calculate real-space properties: onsite parameters, partial densities of 
states, and local charges. In impurity atoms that are empirically known to
be embrittler (N and O) we found that the $2s$ orbital is bound to 
the impurity and therefore does not participate in the covalent bonding. 
In contrast, the corresponding $2s$ orbital is found to be delocalized 
in the cohesion enhancers (B and C). Each of these impurity atoms is found 
to acquire a net negative local charge in NiAl irrespective of whether
they sit in the Ni or Al site. The embrittler therefore reduces
the total number of electrons available for covalent bonding by removing 
some of the electrons from the neighboring Ni or Al atoms and localizing 
them at the impurity site. We show that these correlations also hold
for silicon, phosporus, and sulfur.
\end{abstract}
\pacs{PACS number(s): not available}]

\section{Introduction}

The development of better nickel-based superalloys has paced the 
construction of larger, more powerful, and more fuel efficient aircraft 
and industrial gas turbines.\cite{stoloff1996,darolia1991} Aluminum is the most
important alloying element in nickel, for both strength and oxidation
resistance.\cite{stoloff1996,blankenship1996,miracle1993}
Pure NiAl, which crystallizes in the B2 structure, has low density, high 
melting temperature of 1638$^\circ$C (melting temperature for the fcc nickel
is 1455$^\circ$C), and good
electrical and thermal conductivity.\cite{miracle1993,miracle1995}
Its practical application, however, is limited by poor toughness and damage
tolerance at room temperature\cite{miracle1995} and brittle grain-boundary
fracture at ambient and elevated temperature.\cite{liu1997}
The strength and other properties of NiAl can be modified by adding 
various impurity atoms. Typical modern nickel-base superalloys contain
eight or more different elements, each with specific functions with respect
to strength, alloy stability, and environmental resistance.\cite{stoloff1996}
Certain elements have been found to be deleterious to the properties of NiAl,
among them are nitrogen, oxygen, silicon, phosphorus, and sulfur.\cite{stoloff1996} 
The presence of these elements must be controlled during the melting processes. 
On the other hand, some other elements are desirable because they improve the 
cohesive properties of NiAl. Chromium impurities are important for improving 
its hot corrosion resistance, while boron, carbon, and zirconium provide
improved resistance of grain boundaries to fracture at elevated temperatures.
\cite{stoloff1996,liu1997} Stoloff has given an extensive list of atomic additives 
and their effects on the properties of nickel-base superalloys.\cite{stoloff1996}


Boron is the main grain-boundary strengthener in NiAl. The beneficial effect of
boron additives was first observed by Aoki and Izumi in 1979.\cite{aoki1979}
Boron has a strong tendency to segregate to grain boundaries and it can improve 
the tensile ductility of a polycrystal by an order of magnitude.
\cite{liu1997} This increase in tensile ductility is accompanied by a change
in the fracture mode from brittle intergranular to ductile transgranular
\cite{liu1997} which clearly shows the effectiveness of boron in improving 
the intergranular cohesion in a polycrystal. The strengthening effect of boron
additives has also been observed when they are present as impurity atoms in bulk.
The improvement in strength obtains even with a small concentration of boron
dopants: 30 weight ppm of boron can give rise to a 30\% increase in yield strength.
\cite{george1990} In addition to boron, carbon, which is the element next to boron 
in the periodic table, is also a potent strengthener in NiAl.\cite{miracle1995} 


In contrast to boron and carbon, oxygen and nitrogen are known to be harmful
to the cohesion in NiAl. Indeed oxidation is among the most common degradation
mechanisms in many metals and alloys.\cite{kubaschewski1962,doychak1995} 
In NiAl, oxygen will selectively attack the least noble constituent, which is
aluminum, and form the stable oxide product ${\rm Al}_2{\rm O}_3$.\cite{aitken1966}
The rate of formation of NiO is negligible 
compared to that of ${\rm Al}_2{\rm O}_3$.\cite{miracle1993,liu1997,aitken1966}
This strongly-preferential bonding has also been shown to occur in some recent 
first-principles calculations\cite{lozovoi2000,djajaputra2001} and it may be 
among the key microscopic ingredients for the formation of various mesoscopic 
structures (e.g., pores, cracks, and blisters) created during an oxygen 
attack on an intermetallic alloy. \cite{kubaschewski1962} 
In the extreme, oxygen can cause the pesting degradation phenomenon which 
happens when some polycrystalline samples are 
heated in air within a certain range of intermediate to high temperatures.
\cite{doychak1995,aitken1966} This process, which is 
essentially a spontaneous disintegration of the polycrystalline alloy 
to powder, can take place in a matter of several hours.
\cite{chuang1990,westbrook1964}


\begin{figure}
      \epsfxsize=50mm
      \centerline{\epsffile{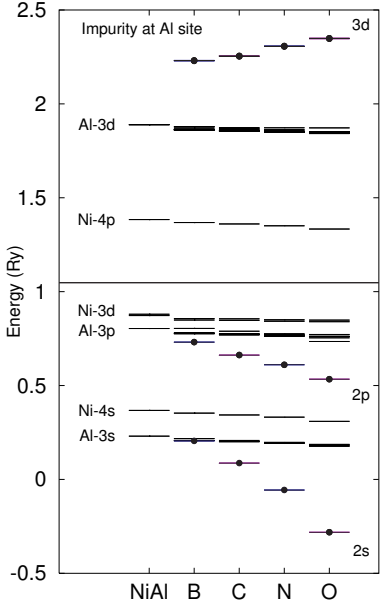}}
\smallskip
\caption{Onsite parameters for orbitals in 16-atom supercells of NiAl 
with one impurity atom substituting for Al.
The leftmost column gives the onsite parameters for pure NiAl. The lines
with dots are the onsite parameters for the $2s$, $2p$, and $3d$ orbitals 
at the impurity atom. The horizontal line at $E_F = 1.0475$ Ry 
is the Fermi level of the pure NiAl. The Fermi energy for the supercell
is 1.0105, 1.0085, 0.9943, and 0.9927 Ry for B, C, N, and O, respectively.} 
\label{XAl7Ni8_levels}
\end{figure}

\begin{figure}
      \epsfxsize=60mm
      \centerline{\epsffile{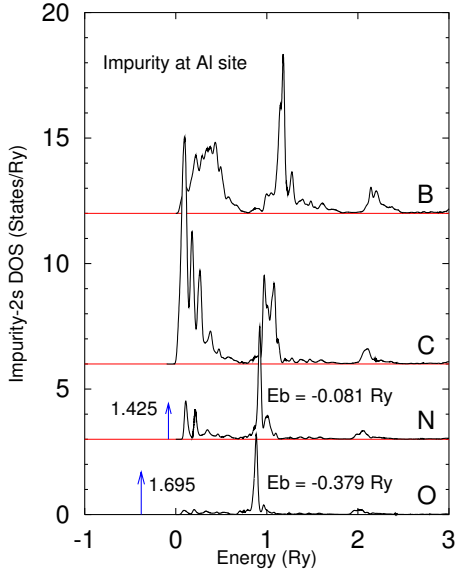}}
\smallskip
\caption{Site-projected $2s$ density of states at the impurity atom for
the case where it is substituting an Al atom. Notice the strong 
resonance at the bottom of the band in the case of carbon. The delta
functions shown in the spectra for N and O are actually very narrow bands
(in our supercell calculation) with bandwidth of 0.0226 Ry and 0.00198 Ry, 
respectively. The numbers next to the arrows are the partial weights 
of the impurity $2s$ state in the band, while $E_b$ denotes the center
of the band. The Fermi level of NiAl is at $E_F = 1.0475$ Ry. 
Each spectrum has been given a separate vertical shift for clarity.} 
\label{XAl7Ni8_DOS}
\end{figure}

There have been several first-principles calculations in the literature on
the effects of impurities on the cohesion in nickel aluminides and related
alloys. Sun {\it et al.} have studied the effects of boron and hydrogen on 
Ni$_3$Al using a full-potential linear muffin-tin orbital (FPLMTO) method.
\cite{sun1995} They emphasized the increase of the interstitial bonding charge
as the origin of the beneficial effect of boron. Wu {\it et al.} calculated
the effects of boron and phosphorus on the grain-boundary cohesion of iron
using a full-potential linear augmented plane-wave (FPLAPW) method.
\cite{wu1994} They showed that a combination method of the thermodynamic
theory of Rice and Wang\cite{rice1989,anderson1990} and first-principles 
total-energy calculations can be used to determine the grain-boundary 
embrittlement potency of a given impurity. Using the same combination 
method they have also studied the effects of hydrogen and carbon impurities
in iron and hydrogen, boron, and phosporus in nickel.
\cite{wu1996,geng1999,zhong2000} Vacancies and antistructure defects
in transition-metal aluminides have been studied by several different 
groups.\cite{fu1995,bester1999,rasamny2001}

Previous first-principles studies on the effects of impurity atoms 
in nickel aluminides have generally focused on, and drawn their conclusions 
from, the calculated total energy and electronic charge densities.
Insights into the bonding and hybridization in the system, 
however, can usually be obtained more clearly by working with 
localized basis functions and using the simpler tight-binding 
representation.\cite{colinet1989,muller1998,pankhurst2001}
Recently we have shown that accurate tight-binding parameters can
be obtained directly from the FPLMTO method.\cite{djajaputra2002} 
In this paper we have used this method to perform a systematic study 
of impurities on NiAl. The motivation for carrying out a systematic 
study is the widely different effects that can be caused by ``nearby''
atoms in the periodic table. It is not obvious, e.g., why, along the 
$2p$ row, boron and carbon are good cohesion enhancers in NiAl while 
the next elements, nitrogen and oxygen, are embrittlers. The present
study has been carried out in an effort to find the answer to this
question. In the next section, we will give a brief description of the
FPLMTO method that we use. The rest of the paper presents the results
of our calculations.

\section{FPLMTO Method}


\begin{figure}
      \epsfxsize=50mm
      \centerline{\epsffile{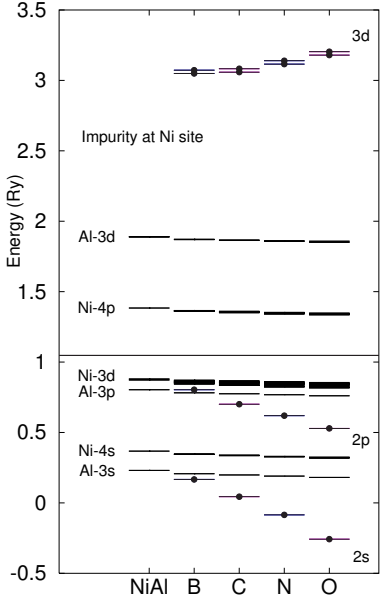}}
\smallskip
\caption{Onsite parameters for orbitals in 16-atom supercells of NiAl 
with one impurity atom substituting for Ni.
The leftmost column gives the onsite parameters for pure NiAl. The lines
with dots are the onsite parameters for the $2s$, $2p$, and $3d$ orbitals 
at the impurity atom. The horizontal line at $E_F = 1.0475$ Ry 
is the Fermi level of the pure NiAl. The Fermi energy for the supercell
is 1.0283, 1.0185, 1.0051, and 1.0014 Ry for B, C, N, and O, respectively.} 
\label{XNi7Al8_levels}
\end{figure}

\begin{figure}
      \epsfxsize=60mm
      \centerline{\epsffile{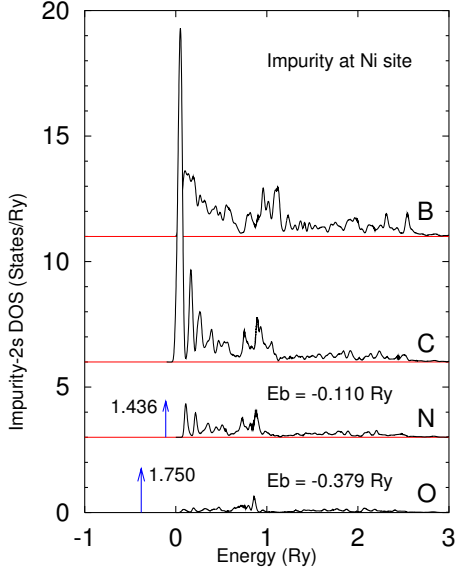}}
\smallskip
\caption{Site-projected $2s$ density of states at the impurity atom for
the case where it is substituting a Ni atom. Notice the strong 
resonance at the bottom of the band in the case of carbon. The delta
functions shown in the spectra for N and O are actually very narrow bands
(in our supercell calculation) with bandwidth of 0.0225 Ry and 0.00158 Ry, 
respectively. The numbers next to the arrows are the partial weights 
of the impurity $2s$ state in the band, while $E_b$ denotes the center
of the band. The Fermi level of NiAl is at $E_F = 1.0475$ Ry.
Each spectrum has been given a separate vertical shift for clarity.} 
\label{XNi7Al8_DOS}
\end{figure}

We use the Wills-Price all-electron full-potential implementation 
of the LMTO method.\cite{wills2000,price1989,price1992} In FPLMTO, 
no assumption is made about the form of the wave functions, 
charge density, or potential.
The muffin-tin potential is used only to construct the LMTO basis functions
but the final wave functions, and other quantities derived from 
them, are not limited to such form.\cite{djajaputra2002} Relativistic
Dirac equations are used for the core states, while the valence states 
are treated semirelativistically without spin-orbit coupling. For the
exchange-correlation potential, we use the parametrization of Vosko, 
Wilk, and Nusair.\cite{vosko1980} Within the muffin-tin spheres, 
lattice harmonics with angular momentum $l \leq 8$ are used. NiAl 
is a good paramagnetic metal (it has no measurable magnetic ordering down 
to temperatures of a few
Kelvin\cite{willhite1973,kulikov1999}) therefore we do not use 
spin polarization in our calculation.

Impurity is incorporated in our FPLMTO calculations by using a 16-atom
supercell.\cite{djajaputra2001} NiAl crystallizes in B2 structure which 
is a bcc-based structure with one atom (Ni or Al) occupying the center 
of the cube $({1 \over 2},{1 \over 2},{1 \over 2})$ and the other 
(Al or Ni) at the corner of the cube $(0,0,0)$. The cubic supercell is 
constructed from $2^3$ NiAl unit cells and the impurity atom is placed
at the center of the supercell. Each atom is assigned a minimal basis
set consisting of 9 $(spd)$ orbitals. Since we want to work with 
localized orbitals, the interstitial parameter for each orbital has
been uniformly set at $\kappa = -0.2$ a.u.
This gives well-localized FPLMTO basis functions with an 
envelope that decays roughly as $\exp({-|\kappa| r})$.\cite{djajaputra2002}

The standard FPLMTO method self-consistently calculates the basis 
functions, along with the corresponding charge density and the resulting
total energy, by working in momentum space. The program computes the
matrix elements of the hamiltonian, $H_{\alpha \beta}(\boldk)$, and 
the overlap, $S_{\alpha \beta}(\boldk)$, matrices from which the 
energy bands $\varepsilon(\boldk)$ are obtained by diagonalization. 
From these matrix elements in momentum space, we have calculated the
matrix elements in real space by direct Fourier transform:
   
\begin{equation}
H_{\alpha \beta}(\boldk) = \sum_j \exp{(i \boldk \cdot {\bf R}_j)} 
H_{\alpha \beta}({\bf R}_j).
\end{equation}

\noindent The onsite parameters are simply the hamiltonian matrix elements, 
in real space, between identical orbitals, $H_{\alpha \alpha}(\boldR = {\bf 0})$. 
This is computed by using an additional code built on top of our 
FPLMTO program. The distribution of onsite parameters is an important 
ingredient in, e.g., Anderson's theory of diagonal localization.
\cite{anderson1958,ziman1979,mott1990,tong1980} In this theory, the
distribution of onsite parameters, characterized by the width of the 
distribution $W$, competes with the strength of the hybridization between
the orbitals, which in the impurity case can be taken to be the bandwidth 
of the parent system $B$. Electron localization is more favorable for 
large values of $W/B$.\cite{ziman1979}

To obtain a measure of the hybridization strength between the orbitals in
the system, without having to deal explicitly with the multiplicity 
of hopping and overlap parameters, one can instead examine the density 
of states (DOS) and its atomic-site and angular-momentum projections.
\cite{varma1980,wilson1980,dunaevskii1980}
Spin projection is unnecessary since NiAl is paramagnetic and we do 
not use spin polarization in our calculations. In this paper the 
total DOS and its projections have been computed using the standard
tetrahedron method with 35 points in the irreducible wedge
of the cubic Brillouin zone. The total DOS
is calculated by summing the contributions from all bands and all
tetrahedra:\cite{gilat1976}

\begin{equation}
\rho(E) = \sum_{n,\boldk_c} g_n(\boldk_c; E),
\end{equation}  

\noindent where $n$ is the band index while $\boldk_c$ is the index 
for the tetrahedra. The site (index $i$) and angular-momentum (index $l$) 
projection of the DOS is obtained by multiplying each contribution 
with its decomposition weight $w_{nil}(\boldk_c)$ which is obtained 
from the wavefunctions:

\begin{equation}
\rho_{il}(E) = \sum_{n,\boldk_c} w_{nil}(\boldk_c) \cdot g_n(\boldk_c; E).
\end{equation}  

Standard DOS decomposition in the FPLMTO method differentiates between 
the muffin-tin (MT) and the interstitial components of the electron
distribution.\cite{sun1995} Further site and angular-momentum 
decomposition, i.e. the calculation of the weights $w_{nil}(\boldk_c)$,
is then performed {\it only} on the part of the LMTO wavefunction
{\it inside } the MT spheres. The interstitial part is not considered to 
belong to any particular site and therefore is not subjected to further
decomposition. It should be noted that this differentiation between
MT and interstitial charge is an artificial one since it depends on 
the size of the MT sphere which, in common practice, is set 
rather arbitrarily by the user of the FPLMTO code. Furthermore, the 
interpretation of such a decomposition is difficult since, e.g.,
the integrated spectral weight for a particular atom ($v_i$) is, in
general, less than the total number of valence electrons assigned
to it ($n_i$):

\begin{figure}
\epsfxsize=60mm
\centerline{\epsffile{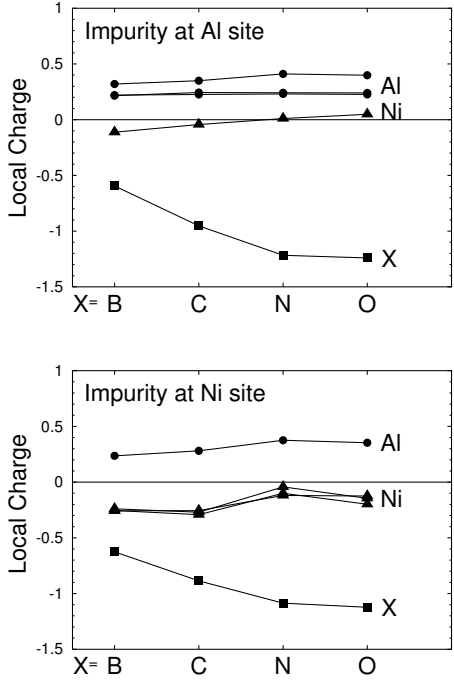}}
\smallskip
\caption{Local charge (in units of the electron charge $|e|$) 
induced on each atom in the supercell for the case 
where an impurity atom (X = B, C, N, and O) is substitutionally placed at 
an Al site (top panel) and a Ni site (bottom panel). Aluminum charges 
are marked by filled circles ($\bullet$), Ni by filled triangles, 
while the impurity charges are shown as 
filled squares. In pure NiAl, Al has a charge of $+0.2$ while,
from charge neutrality, Ni has the opposite charge of $-0.2$.}
\label{LocalCharges}
\end{figure}

\begin{equation}
v_i^{\rm (LMTO)} = \sum_{l=0}^{l_m} \int_{-\infty}^\infty 
\rho_{il}^{\rm (LMTO)}(E) \ dE \ \leq \ n_i.  
\end{equation}

\noindent Note that the summation over the angular momenta extends up to
$l_m$, which is a free parameter in an FPLMTO calculation (this parameter
is set to 8 in this work). In general, this parameter is different  
from (usually much greater than) the highest angular momentum $L_m$ that 
one uses in defining the FPLMTO basis functions ($L_m = 2$ for $spd$ basis
that we use here). Inside each MT, tails from the basis functions centered 
at other MTs give rise to higher angular-momentum harmonics when 
expanded relative to the center of the MT sphere. The parameter $l_m$ is 
the cutoff value used in this expansion.\cite{djajaputra2002}

Instead of using this MT decomposition, in this paper we have chosen 
to use an orthogonal decomposition which is the one used in 
tight-binding systems. The FPLMTO non-orthogonal matrix elements,
$H_{\alpha \beta}(\boldk)$ and $S_{\alpha \beta}(\boldk)$, are first
transformed into an orthogonal system by L\"owdin transformation.
\cite{lowdin1956} Since this is a symmetry transformation which 
does not mix components of different angular momenta,
\cite{slater1954,altmann1995} the weights for the $l$-projected 
DOS can be obtained readily from the resulting L\"owdin eigenvectors.
Details on this scheme have been presented in an 
earlier paper.\cite{djajaputra2002}  
In this decomposition, the angular momentum expansion extends 
only to $L_m$ and the total atomic weight is equal to the number
of the assigned valence electrons since the interstitial
continuation of each FPLMTO basis function has been incorporated
properly:

\begin{equation}
v_i^{\rm (TB)} = \sum_{l=0}^{L_m} \int_{-\infty}^\infty
\rho_{il}^{\rm (TB)} (E) \ dE = n_i.
\end{equation} 

\noindent This decomposition method is more appropriate to use in our
case since we exclusively use localized FPLMTO basis functions
(specified by negative $\kappa$ parameter). It should be pointed
out that the L\"owdin transformation to orthogonal 
system is used solely to obtain the decomposition weight 
$w_{nil}(\boldk_c)$ for the local DOS; elsewhere in this paper 
we work directly with {\it non}-orthogonal TB systems.
The onsite parameters displayed in Figs.~\ref{XAl7Ni8_levels}
and \ref{XNi7Al8_levels}, e.g., are matrix elements of the 
hamiltonian operator in the original non-orthogonal FPLMTO basis;
they are not, and should not be confused with, the matrix elements in a 
L\"owdin orthogonal basis which are nowhere presented or analyzed 
in this paper.

From the projected DOS, the total number of electrons residing 
on each atomic site can be obtained by integrating the 
corresponding DOS up to the Fermi energy:

\begin{equation}
q_i = \sum_{l=0}^{L_m} \int_{-\infty}^{E_F}
\rho_{il} (E) \ dE.
\label{LocalChargeFormula}
\end{equation}

\noindent Here $E_F$ is the self-consistent Fermi energy calculated
for each supercell (with impurity atom) and not the Fermi energy
of the pure NiAl system. In the next section, we present the results 
of our calculations for the onsite parameters, projected DOS, 
and the local charges.
 
\section{Computational Results}

\subsection{Boron, Carbon, Nitrogen, and Oxygen}

\begin{figure}
      \epsfxsize=78mm
      \centerline{\epsffile{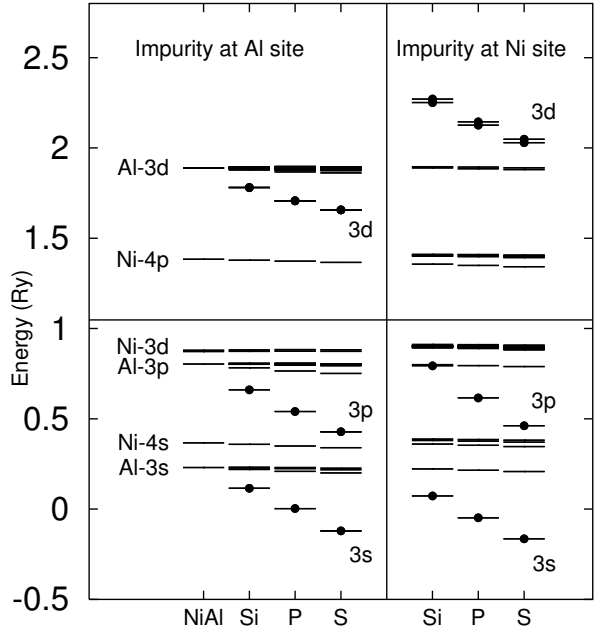}}
\smallskip
\caption{Onsite parameters for orbitals in 16-atom supercells of NiAl 
with one impurity atom (Si, P, or S) substituting for Al (on the left side 
of the vertical midline); and with the impurity substituting for Ni (right side).
The leftmost column gives the onsite parameters for pure NiAl. The lines
with dots are the onsite parameters for the $3s$, $3p$, and $3d$ orbitals 
at the impurity atom. The horizontal line at $E_F = 1.0475$ Ry 
is the Fermi level of the pure NiAl. In the case where the impurity is 
occupying an Al site, the Fermi energy for the supercell is 
1.0440, 1.0465, and 1.0426 for Si, P, and S, respectively. For the case
where it is occupying a Ni site, the Fermi energies are 1.0765 Ry (Si), 
1.0795 Ry (P), and 1.0750 Ry (S).} 
\label{SiPS_levels}
\end{figure}

Fig.~\ref{XAl7Ni8_levels} displays the calculated onsite parameters
in pure NiAl and in 16-atom supercells of NiAl with one impurity
atom substituting for Al. We have used the {\it computed} equilibrium
lattice constant for NiAl (5.3451 a.u.) which is within 2\% of the 
experimental value (5.4450 a.u.). The XAl$_7$Ni$_8$ supercell (here
X stands for the impurity atom) is constructed from
$2^3$ NiAl unit cells\cite{djajaputra2001} with the impurity atom
placed at $(0,0,0)$; Al atoms at $({1 \over 2},0,0)$, $({1 \over 2},
{1 \over 2},0)$, $({1 \over 2},{1 \over 2},{1 \over 2})$, and other
equivalent positions obtained by permuting the $x,y,z$ coordinates;
and Ni atoms at $(\pm {1 \over 4},\pm {1 \over 4},\pm {1 \over 4})$
(in units of supercell lattice constant). Note that the positions
of the Ni atoms are all symmetry-equivalent in this supercell. Al atoms,
on the other hand, occupy three inequivalent sites. This gives 
rise to a small splitting of the Al onsite parameters as
can be seen in Fig.~\ref{XAl7Ni8_levels}; the corresponding splitting
of the Ni parameters in the XNi$_7$Al$_8$ supercell can be seen 
in Fig.~\ref{XNi7Al8_levels}. Atomic relaxation has been
shown to produce only a small change in energy\cite{djajaputra2001}
and therefore has been ignored in this work. The small value of
the computed relaxation energy,\cite{djajaputra2001} and the small
size of the onsite-parameter splittings in Figs.~\ref{XAl7Ni8_levels}
and \ref{XNi7Al8_levels}, provide the justification for our neglect 
of atomic relaxation in the present work. It is unlikely that 
relaxation will make large quantitative change in, or rearrange the 
qualitative structure of, the onsite-parameter 
maps in Figs.~\ref{XAl7Ni8_levels} and \ref{XNi7Al8_levels} on
which we will base much of our discussion in this paper.

The utility of plotting the onsite parameters systematically, 
as in Fig.~\ref{XAl7Ni8_levels}, comes from the fact that it shows
clearly how well the $2s$ and $2p$ parameters of boron match
those of the corresponding $3s$ and $3p$ orbitals of aluminum, 
and how rapidly this compatibility deteriorates as we go from 
boron to oxygen. To our knowledge this almost-perfect compatibility 
has never been pointed out previously in the literature.  
The onsite parameters for the $3d$ states of the
impurity atoms are all much higher than the Al-$3d$ 
parameters. Although results from the local density approximation 
(LDA) for the excited states are known in general to be less accurate 
than the corresponding results for the occupied states, we believe 
this visible difference is an important feature in explaining the efficacy 
of boron as a cohesion enhancer in NiAl. The much higher B-$3d$
parameters would allow the delocalized B-$2s$ and B-$2p$ states
to create wider bands centered at their corresponding onsite
parameters which, as we pointed out previously, match closely to 
those of Al-$3s$ and Al-$3p$. The overall cohesion is therefore 
improved by increasing the bond order (the difference in 
occupancy between bonding and antibonding states).\cite{sutton1996}

The bottom of the pure-NiAl bands lies just above the zero energy
in Fig.~\ref{XAl7Ni8_levels}. It can therefore be seen clearly
that the C-$2s$ onsite parameter sits just above this bottom 
while those of N-$2s$ and O-$2s$ orbitals lie below the 
main manifold of pure NiAl. As in standard scattering theory,
\cite{economou1979} this situation opens the possibility for the
existence of resonance or bound states. In Fig.~\ref{XAl7Ni8_DOS}
we show the $2s$ projected DOS at the impurity atoms. The total 
weight under each curve is equal to 2 (due to spin sum) to within
2\% accuracy. For this case, where the impurity atom is occupying
an Al site, the nearest neighbors of the impurity atom are 
Ni atoms. The main feature of the DOS for B-$2s$ is a broad
band which is cleaved by its interaction with the neighboring 
$3d$ orbitals of Ni.\cite{economou1979,castro1993} 
This is markedly different from the DOS for C-$2s$ in which the 
dominating feature is the strong resonant peak at the bottom of the
spectrum. As we move on to N-$2s$ and O-$2s$, the onsite 
parameters for these orbitals are deep enough to localize 
the electrons in a bound state. This results in a transfer of 
the spectral weight from the continuum to the bound state.
In our supercell calculation, the bound state is not manifested
as a true delta function but it rather appears as a very narrow
band (with bandwidth of 23 mRy and 2 mRy for N-$2s$ and
O-$2s$, respectively, for the case where the impurity is placed
at an Al site) which is separated by a gap from the main 
spectrum and is displayed as a vertical arrow in  
Fig.~\ref{XAl7Ni8_DOS}. This narrow band still contains small 
hybridization components from other orbitals (this, of course,
is just an artifact of a supercell calculation) which, as expected, 
diminish as we go from N to O. The total weight of the 
impurity-$2s$ state in the narrow band is displayed next to its arrow
in Fig.~\ref{XAl7Ni8_DOS} while the rest of the weight still
remains spread out thinly in the continuum. 

\begin{figure}
      \epsfxsize=60mm
      \centerline{\epsffile{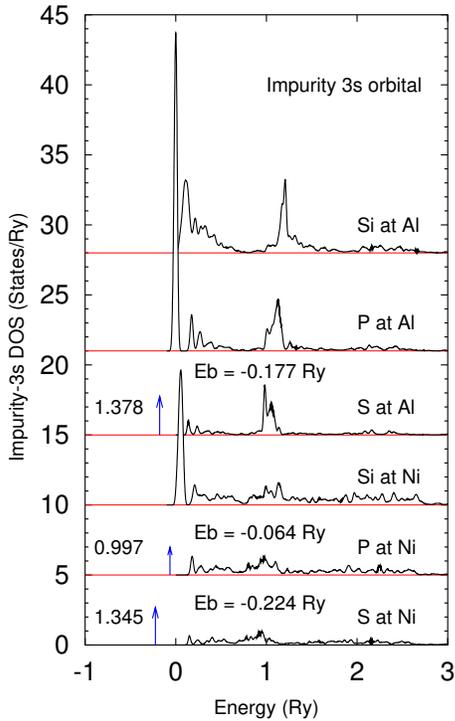}}
\smallskip
\caption{Site-projected $3s$ density of states at the impurity site
(Si, P, or S). A bound state is
formed for the cases of phosporus atom at Ni site and sulfur atom at Al 
or Ni site. The delta functions shown in the spectra are actually very 
narrow bands (in our supercell calculation) with a bandwidth of 47 mRy 
(P at Ni), 26 mRy (S at Al), and 16 mRy (S at Ni). The numbers next to 
the arrows are the partial weights of the impurity $3s$ state in the
corresponding narrow band, while $E_b$ denotes the center
of the band. The Fermi level of NiAl is at $E_F = 1.0475$ Ry. 
Each spectrum has been given a separate vertical shift for clarity.} 
\label{SiPS_DOS}
\end{figure}
 
A very similar map of onsite parameters is obtained in the alternative
case where the impurity atom is substituting for a Ni atom, as
shown in Fig.~\ref{XNi7Al8_levels}. The main difference from 
Fig.~\ref{XAl7Ni8_levels} is the fact that the impurity-$3d$ 
levels are pushed to much higher values in this case
(by about $0.7-0.8$ Ry). This feature is mainly due to the 
smaller size of Ni, compared to Al, which increases the kinetic
energy of the orbital (we use a MT radius of 2.30 a.u.~for Al and
1.85 a.u.~for Ni). Combined with the bond-order argument 
described previously, this also provides a heuristic explanation
on why an oxygen impurity would prefer to occupy a nickel site over an 
aluminum site, a result which was recently obtained from a 
full-fledged FPLMTO calculation.\cite{djajaputra2001}
The calculated impurity-$2s$ DOS for the case of impurity at Ni
site is shown in Fig.~\ref{XNi7Al8_DOS}. The general progression
from B to O is the same as in Fig.~\ref{XAl7Ni8_DOS}: A broad band
for B-$2s$, strong resonance for C-$2s$, and bound state with 
an increasing binding energy for N-$2s$ and O-$2s$. Since
the impurity atom is surrounded by Al nearest neighbors in
this case, instead of Ni atoms, we do not see as strong 
a band cleavage around the energy of the Ni-$3d$ orbitals as 
seen in Fig.~\ref{XAl7Ni8_DOS}. Below the Fermi energy, the 
similarity of the results obtained for impurity at Al and Ni 
sites shows that these features, e.g.~the compatibility of
the onsite parameters for B and Al in NiAl environment, are 
largely independent of the atomic arrangement in the crystal.  
This is not surprising since an onsite parameter is
sensitive only to the average potential at its atomic site.
This suggests that our results in this paper, which 
have been obtained for NiAl host using 16-atom supercells,
may have some relevance also to other nickel-aluminide
alloys with different concentrations of impurity atoms.

Fig.~\ref{LocalCharges} shows the charges induced on each atom 
in the supercell which have been calculated by substracting
the total number of electrons on the site, $q_i$ in 
Eq.(\ref{LocalChargeFormula}),
from the assigned number of valence electrons $n_i$:

\begin{equation}
Q_i = n_i - q_i. 
\end{equation}

\noindent In all cases, Ni is found to be 
more electronegative than Al (the Pauling electronegativity
of Ni and Al is 1.91 and 1.61 respectively\cite{emsley1994}). 
In pure NiAl, Al has a charge of $+0.2$ (in units of electron 
charge $|e|$) while, from charge neutrality, Ni has the opposite 
charge of $-0.2$. The impurities from the $2p$ row that we have 
studied in this work have Pauling electronegativity of
2.04, 2.55, 3.04, and 3.44 for B, C, N, and O, respectively.
\cite{emsley1994} It can be seen that this electronegativity 
trend is followed rather well in Fig.~\ref{LocalCharges}.
In the case of impurity atom at Al site (top panel in 
Fig.~\ref{LocalCharges}), N and O are sufficiently 
electronegative to change the sign of the induced charge 
on their Ni nearest neighbors, relative to the sign of the 
corresponding charge when B or C is present. Thus a 
portion of the valence electrons localized at the N 
or O bound state comes from their nearest-neighbor Ni atoms.
In the alternative case where the impurity is occupying the
Ni site, a jump in the induced charge on the Ni atoms is 
clearly seen in the bottom panel of Fig.~\ref{LocalCharges}.
Although in this case they are no longer the nearest neighbors
of the impurity (since they are separated from it by the 
Al atoms), the formation of the bound state in N and 
O still has a substantial effect on the Ni atoms. 
Two reasons may be given to explain this strong interaction
between the impurity and the Ni atoms. First, the Al nearest
neighbors are already positively charged, therefore it is 
relatively harder for the impurity atom to attract their
electrons. Second, the DOS of NiAl is dominated by strong 
Ni-$3d$ peaks which are situated just below 
the Fermi energy.\cite{djajaputra2001}
These peaks are sufficiently wide to suggest that the Ni-$3d$ 
electrons in this alloy are well delocalized. Their
proximity to the Fermi energy then strongly expose them 
to changes in the potential as that caused by the formation of
a bound state on a nearby atom.

\subsection{Silicon, Phosphorus, and Sulfur}

The elements from the $3p$ row of the periodic table: Si, P, 
and S, have been known to be strong embrittlers in NiAl.
\cite{stoloff1996} It is therefore interesting to examine
whether the correlation that we have obtained in the previous
subsection between the matching of the onsite parameters
and the macroscopic embrittling/strengthening potency 
of the impurity persists also for these elements. 
Fig.~\ref{SiPS_levels} displays the calculated onsite 
parameters for orbitals in 16-atom supercells of
NiAl containing one impurity atom (Si, P, or S) which substitutes
for an Al (shown on the left side of the vertical midline in 
Fig.~\ref{SiPS_levels}) or a Ni atom (shown on the right side
of the midline). A major difference from the corresponding 
plots of onsite parameters in Fig.~\ref{XAl7Ni8_levels} 
and Fig.~\ref{XNi7Al8_levels} is the decreasing trend of 
the impurity-$3d$ levels as we go to higher atomic number 
(from Si to S). This is due to the fact that the basis 
orbitals that we use in this case ($3s$, $3p$, and $3d$) 
all have the same principal quantum number. The difference
in their levels therefore originates mainly from the 
difference in the effective centrifugal potential 
(the $l(l+1)r^{-2}$ term in the radial \Schrodinger equation),
which is independent of the atomic number.
\cite{woodgate1989,flynn1972}
In contrast, the basis orbitals that we use for the $2p$
elements in Fig.~\ref{XAl7Ni8_levels} and Fig.~\ref{XNi7Al8_levels}
($2s$, $2p$, and $3d$) come from two different principal
quantum number shells. In this case, in addition to the 
centrifugal potential, the splitting among the onsite levels is 
also determined by the Coulomb potential of the nucleus 
which increases with the atomic number. Thus the $2p$ level
decreases in concert with the $2s$ level while the splitting
between them and the $3d$ level increases with the atomic number
as we go from boron to oxygen in Figs.~\ref{XAl7Ni8_levels}
and \ref{XNi7Al8_levels}.

Fig.~\ref{SiPS_DOS} shows the resulting projected DOS for the
lowest-lying valence ($3s$) state of Si, P, and S at 
the impurity site. As in Fig.~\ref{XAl7Ni8_DOS}, when the 
impurity is placed at the Al site, its DOS features a peak close
to the Fermi energy due to its strong hybridization with the 
$3d$ states of its neighboring Ni atoms. 
Except for the case of Si at Al site, 
where the resonance at the bottom of the spectrum is relatively weak,
the DOS curves in Fig.~\ref{SiPS_DOS} are all dominated 
either by a very strong resonance
(P at Al; Si at Ni) or a bound state that is completely
separated from the main spectrum (S at Al; P at Ni; and S at Ni).
Silicon, phosporus, and sulfur are known to be embrittlers in
NiAl.\cite{stoloff1996} These results therefore support the
correlation that we have obtained in the previous section that 
relates the localization of the valence electrons at the 
impurity site with the macroscopic embrittling character of
the impurity atom when it is present in NiAl.  
 
Although the weak resonance in the case of Si at Al site 
seems to defy this correlation (note that, for 
reason of presentation clarity, the projected-DOS curves that 
we show in Figs.~\ref{XAl7Ni8_DOS}, \ref{XNi7Al8_DOS}, and 
\ref{SiPS_DOS} have been obtained by convoluting the FPLMTO
DOS with a Gaussian smearing function of width about 10 mRy), it
should also be noted that its 
$3d$-state level in Fig.~\ref{SiPS_levels} is much lower 
than the corresponding $3d$ level for, e.g., boron or 
carbon in Fig.~\ref{XAl7Ni8_levels}. As has been pointed out in 
the previous subsection, this much-lower $3d$ level exerts an
`onsite pressure' on its lower-lying $s$ and $p$ states against
forming a wider band (due to its orthogonality with these states).
This results in narrower bands under the
Fermi level and, consequently, in reduced bond order and weaker
metallic character of bonding around the impurity site. This 
may explain why silicon is an embrittler in NiAl while carbon, 
which has a similar set of onsite parameters below the Fermi 
energy as can be seen by comparing Fig.~\ref{XAl7Ni8_levels} and
Fig.~\ref{SiPS_levels}, is in contrast a cohesion enhancer.

\section{Summary}

In this paper we have performed a systematic study of impurity
hybridization in the refractory alloy NiAl. Impurity atoms from
the $2p$ row (B, C, N, and O) and the $3p$ row (Si, P, and S) 
of the periodic table have been examined. The purpose of this 
study is to understand the origin of the embrittling/strengthening 
property of impurity atoms in alloys in terms of the 
compatibility of their onsite parameters and their orbital 
hybridization. We found that the onsite 
parameters of boron, which is the prime cohesion enhancer in NiAl,
are highly compatible with those of the NiAl host below the Fermi
energy. In addition, its higher-lying atomic levels are located
higher than the corresponding levels for Al. This allows the
$2s$ and $2p$ states of boron to hybridize more strongly with 
the orbitals at the neighboring atoms, form wider valence bands
centered below the Fermi energy, and increase the bond order. 
These two properties, the 
compatibility of the onsite parameters and the relative location 
of the higher-lying states of the impurity atom, have been found 
useful in understanding the electronic structure of the impurities
and their effects on the cohesion in NiAl.

It is a pleasure to thank Dr. Leonid Muratov for many 
stimulating discussions and useful inputs.
This work was supported by AF-OSR Grant No. F49620-99-1-0274.

\end{document}